# Managing Smartphones Signalling Load in UMTS Networks: A Practical Analysis

Ayman Elnashar, Mohamed A. El-saidny, Mohamed Reda

## Abstract

*Smartphone users and application behaviors add high pressure to apply smart techniques that stabilize the network capacity and consequently improve the end-user experience. The massive increase in smartphone penetration engenders signalling load, which exceeds the network capacity in terms of signaling. The signalling load leads to network congestion, degradation in the network KPIs. The classical way to tackle the signalling is by network expansion. However, this approach is not efficient in terms of capital expenditure (CAPEX) and also in terms of efficient utilization of the network resources. More specifically, the signaling domain becomes overloaded while the data domain are underutilized. In this paper, two UMTS air-interface features; Cell-PCH (paging channel) and enhanced fast dormancy (E-FD) are analyzed to mitigate the signalling load. Practical performance analysis is conducted based on results from commercial UMTS networks. The deployment of these features offers major improvement in network KPIs and significant relief in the signaling load. It is concluded that these two features in addition to several optimization techniques, discussed in this paper, provide solution to the smartphone signaling load and efficiently utilize the available network and spectrum resources while providing users with better always-on connectivity and improved battery consumption.*

## I. Introduction

The smartphone users and their applications in a network put disproportionate burden on networks, caused by their typical behavior of many connections and low data volume transfers per connection. Even with a small volume of smartphone in a network, they still produce a major impact on the PS (packet switch) domain. Figure 1 illustrates one practical example from a commercial network with low smartphone penetration of 13% generating 82% of the total PS call attempts. This behavior demonstrates how smartphones are consequently contributing to a significant amount of PS signaling load compared with other devices.

With the massive increase in smartphones penetration, the signaling load is expected to exponentially increase which overloads the control plane of the UMTS network interfaces [1]. The following are the main causes of the smartphones signaling load [1], [2], [3]:

a) Most smartphones automatically attach to the network and activate a PDP context in order to access the Internet and obtain up-to-date information anywhere and anytime. As a result, almost each smartphone subscriber has a PDP context and the average activation ratio is much higher, which places high pressure on the UMTS network capacity as smartphone penetration increases.

b) Smartphones consume large amount of power due to their large screen, long online time, and diversified mobile applications. To save power, smartphones have adopted fast dormancy technology. If there is no data transmission within a short time (usually 3 to 10 seconds), smartphones will automatically abort the wireless connection and switch to the idle mode. The connection will be reestablished when the data transmission is needed. All these generate large amounts of signaling load.

c) A smartphone makes its best to be always online. If the activation fails due to breakdown of network node, no service subscription, or insufficient capacity, the smartphone will repeatedly try to activate PDP context. As a result, the activation signaling on the network increases sharply and the network congestion or overload occurs consequently.

d) Side effects of mobile malware, subscribers with high frequency communication sessions, poorly designed mobile applications and unwanted traffic from Internet hosts outside the mobile network (see [2] and [3] and references therein).

The UMTS networks are not typically designed to manage such excessive signaling load. The frequent access of applications, such as social media sites, stimulates a signaling load that impact many network elements. Also, many over-the-top (OTT) applications with frequent synchronization with host servers via the operator's network can cause signaling spikes and accordingly network congestion and outage especially in major events. This has motivated the industry to promote best practices for developing "network-friendly" mobile applications [4], [5],

[2]. The analysis in [1] and [2] are based on specific signaling storm due to deliberate malicious activity that aims to disrupt mobile services. On the other side, the analyses in [6]-[10] have focused on analyzing signalling behavior from an energy consumption perspective.

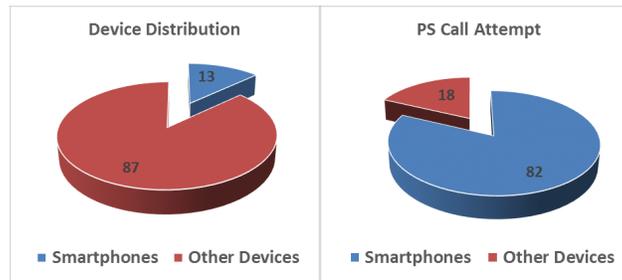

Figure 1: Smartphone distribuation versus PS call attempts

As a result of the excessive signalling load, the network resources are inefficiently utilized with the increasing usage of smartphone applications and data connections. Therefore, several features and techniques have been actively adopted by the smartphone suppliers and network vendors [3]-[5]. The adopted solutions can be categorized into three areas: Application layer mitigation approach, network dimensioning and expansion, and deployment of 3GPP smartphone features. The application layer mitigation approach aims at an intelligent traffic management and bundling of delay tolerant applications. This approach is referred to as smart gating where the User Equipment (UE) synchronizes and bundles data activity into batches when the UE is transferring PS data session in background mode without impacting the user experience. The basic function of the gate is to open periodically or whenever the device is in active mode (e.g. device is paged, screen or microphone/speaker is on, or any other user activity). One of these application layer solutions is Qualcomm's Network Socket Request Manager (NSRM) [11]. These types of solutions can be efficient in reducing the signaling overhead, but they are subject to device support and the lack of 3GPP standardization.

On the other hand, network operators can still pursue the direction of re-dimensioning the network by means of adding more 3G carriers (i.e. frequencies within the same band or different bands), increase the baseband resources in the Node-B, and add more signalling processing boards in the radio network controller (RNC) or even by adding more RNCs. However, the main drawback of such approach is the increase of capital expenditure (CAPEX) investment. The addition of multiple 3G carriers may generate excessive signaling because of the need to handle more handovers at the carriers' boundaries as well as the need to handle traffic separation between circuit switch (CS) services (i.e. voice calls) and packet switch (PS) service on each carrier. Moreover, adding more RNCs can produce other planning issues related to re-dimensioning of the RNCs leading to borders overlapping and as a result more signaling load generated by location area update, paging and RNC boundary handling. In summary, this approach lead to signaling capacity expansion while the data capacity may still not be fully utilized and therefore the CAPEX investment is not fully justified.

The main solution being actively considered by network operators is to utilize the advanced 3GPP smartphone features which is the main focus of this paper. 3GPP has an evolved roadmap to improve end-user experience and connectivity. 3GPP has introduced several performance enhancement features as part of the HSPA+ (evolved High Speed Packet Access) roadmap: Enhanced Fast Dormancy (E-FD), Continuous Packet Connectivity (CPC), or Enhanced Cell-FACH (Forward Access Channel) with DRX (Discontinuous Reception). These features provide different approaches that mainly lead to reduction in signaling overhead, saving uplink and downlink resource, RF power reduction, and battery backup time extension. Each solution is subjected to availability of devices and network vendor support. Operators typically consider deploying each feature based on the maturity of devices and network vendor readiness. Some of these features have already proved to be working well from 3GPP further releases, such as 4G LTE (Long Term Evolution). For example, LTE has introduced the C-DRX (Connected DRX) feature that works similar to the CPC in the UMTS and it offers significant gains to the battery and network resources [12].

Some of these features have strong dependencies on older 3GPP features. For example, E-FD is always expected to work better in a network deployed with connected mode states like Cell-PCH (Paging Channel) or

URA-PCH (UTRAN Registration Area Paging Channel). In addition, CPC is expected to work in conjugate with Signaling mapped into HSDPA/HSUPA instead of R99, referred to as SRB over HSPA. This kind of dependencies requires the operators to validate several older features before deploying HSPA+ related ones [13].

The contribution in this paper is threefold. Firstly, we conduct smartphone profiling based on a commercial network analysis to understand smartphones behavior. The relation between smartphone penetration and signaling load is demonstrated. Secondly, we analyze the mitigation solutions from network side to manage and reduce smartphones signalling load in UMTS network. Specifically, we analyze the performance of Cell-PCH and network controlled enhanced fast dormancy (E-FD) and other related techniques. The best practice and deployment results are presented based on practical network results. Thirdly, we present key optimization techniques for Cell-FACH and Cell-PCH states to tackle the impact of deploying the signaling load mitigation techniques.

The paper is organized as follows: section II provides detailed description of radio resource control (RRC) state machine. Also, the Cell-PCH state and its performance results are provided in section II. Smartphones profiling and signaling load is analyzed in section III. The fast dormancy technique is presented in section IV. The performance results of the E-FD is presented in section V. Optimization of Cell-FACH and Cell-PCH states are presented in part VI. Conclusions and future work are summarized in section VII.

## II. RRC State Machine

The RRC layer is part of the overall UMTS protocol, referred to as Layer-3. RRC constitutes the air interface protocol for the control plane signaling messages. In general, signaling messages are needed to regulate the UE behavior in order to comply with the network procedures. Each signaling message from the UE to the network, or vice versa, consists of a set of system parameters. For example, the Node-B needs to communicate the parameters related to mobility procedures as of when the UE needs to handover from one cell to another. These parameters are broadcast to the UE in a specific RRC message. Additionally, the core network signaling is also carried by a dedicated RRC message.

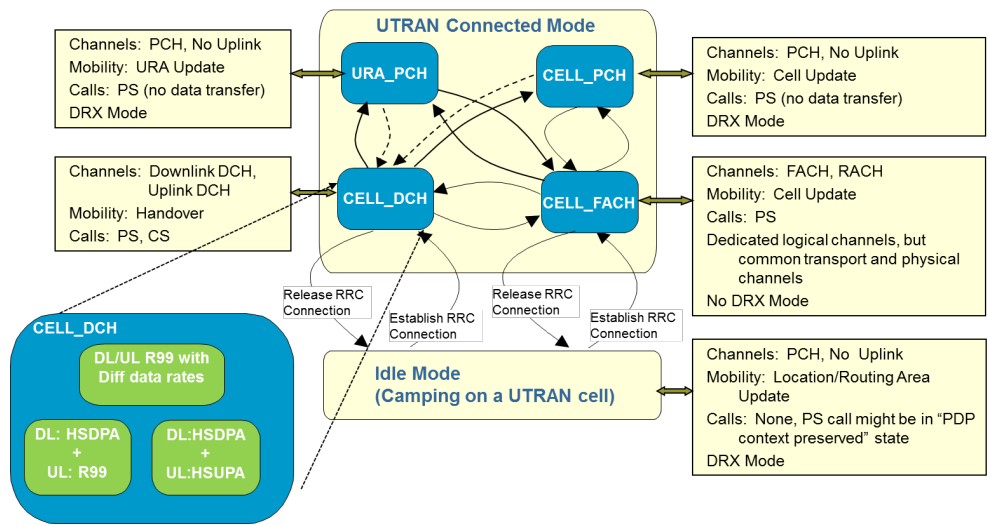

Figure 2: UMTS RRC States and State Transitions

The RRC layer in UMTS is significantly different than the structure of Layer-3 in other systems such as GSM or LTE. 3GPP has designed RRC to function in different modes: Idle and Connected. The connected mode is further categorized into four states: Cell-DCH, Cell-FACH, Cell-PCH and URA-PCH. Each mode and/or state serves the UE depending on the type of service, CS or PS domain services. Figure 2 demonstrates the RRC states and their mobility procedures. It also illustrates the transitions between RRC Idle and Connected modes alongside the different states [14], [15].

*a) RCC Idle Mode*

In the RRC idle mode; there is no radio connection between the UE and the UTRAN for user traffic however the UE can be paged by the core network. The UE shall perform a periodic search for serving cells and monitor the paging messages while it is in this state [14], [15]. This mode is intended to serve the UE whenever there is no voice call (CS domain is idle), or no data traffic (PS domain is idle). A basic signaling connection is needed in this state to serve the UE with System Information Blocks (SIBs) and CS/PS paging messages.

Upon initiating a new call in CS or PS domains, the UE is required to move directly to the RRC connected mode to establish the required service. When a new connection is established, the UE is typically moved to Cell-DCH where a dedicated radio link connection is established with a cell. The Radio Beaer (RB) for the required service is then established and voice or data call can then flow in Cell-DCH state. When the call is released, the UE can be moved to Idle mode again, but depending on being CS or PS call. The Idle state is mandatory in any UMTS network, but expected to be rarely utilized for PS services, especially if Cell-FACH and Cell-PCH states are configured.

*b) RCC Connected Mode*

In the RRC connected mode, the UE and the UTRAN have at most one RRC connection regardless of how many signaling connections exist between the UE and the core network. The signaling messages are only transmitted over signaling radio bearers for either CS or PS domains. The RRC state machine operates as two synchronized peer entities: one is in the UE side, and another is in the UTRAN side in the RNC.

The connected mode consists of Cell-DCH state. This state is the only one in which the UE has a dedicate uplink and downlink connection with the network (i.e. a cell or multiple cells). Voice call can only be carried on this state while PS data packets can be carried on Cell-DCH or Cell-FACH. Cell-DCH is considered as a state where network resources are reserved for the UE. For voice calls, dedicated physical layer channels are established for each UE to carry the voice call. On the other hand, for PS data, the HSPA channel can be shared among multiple users. Therefore, the Cell-DCH state consumes the most of network and UE resources in terms of DL/UL transmit power, respectively, orthogonal variable spreading factor (OVSF) codes, and the device battery.

*1) Cell-FACH State*

As the Cell-DCH state is costly in terms of resource utilization, the 3GPP standard has defined the Cell-FACH state that is typically used to relieve resources. This state is specific to PS domain, to some extend. The physical layer during this RRC state is configured as a common channel instead of a dedicated channel. The users share the resources of the FACH pipe and are scheduled based on data availability. The channel is limited in data rate and accordingly it is not optimal for heavy downloading or uploading activities. While the UE is served in Cell-FACH, it can only connect to one cell i.e., no soft handover and without any sort of closed loop power control. When the PS data buffer increase to a certain predefined threshold, the network move the UE to the Cell-DCH state to be served with high speed data connection such as HSPA(+).

*2) Cell-PCH State*

The drawback of using RRC idle mode is that it requires the UE to always re-connect to the network and initiate a new signaling radio bearer, where signaling load can increase significantly. On the other hand, Cell-PCH and URA-PCH states do not have the same signaling re-connection requirements. These two states are considered the idle mode state inside connected mode where the device battery is efficiently preserved. The mobility aspects in PCH are very much similar to idle. The UE needs only to monitor a cell for a short period in a paging cycle of the paging indicator channel (PICH) in the physical channel. A PDP context with the PS core network is still preserved in these two states [14]. As the PDP context is still retained, a session could be reconnected much faster and with less number of RNC and core network signaling messages. The PDP context dimensioning in core network need to consider such change, as most of the users will remain in the Cell-PCH state once it is enabled. Therefore, the resources of the PS core in term of number of PDP context need to consider such massive increase.

The main difference between Cell-PCH and URA-PCH is in the mobility and planning aspects. Cell-PCH is cell-connected mode while URA-PCH is a UTRAN registration area connected mode. In the cell-connected mode, the location of the UE is maintained by the UTRAN in the accuracy level of a cell. In the URA-connected mode, the UE is only known by the UTRAN in the URA level, a superset of multiple cells. Hence, it is expected that paging overload is more efficient in URA-PCH state, depending on how the URAs are planned.

As explained, it is more efficient for network to utilize PCH states when the data is completely inactive, rather than using the RRC idle mode. Upon an extended time of inactivity, the network can finally move the UE to idle mode and release the PDP context. This timer is usually adjusted to a long period such as 60 min. With a low activity PS connection, the network can utilize Cell-FACH connection, while a transition to Cell-DCH happen when the data activity increases to certain data threshold. A typical value for such threshold is 256 kbps.

Figure 3 illustrates an example of the state transitions between the different RRC states. As shown in this figure, the transition from RRC idle mode to Cell-DCH requires a new establishment of RRC connection with around 15 to 30 RRC messages. On the other side, the transition from Cell-PCH to Cell-FACH requires only 3 RRC messages. The direct transition from Cell-PCH to Cell-DCH, if allowed by the network vendor, it requires around 3 to 12 RRC messages.

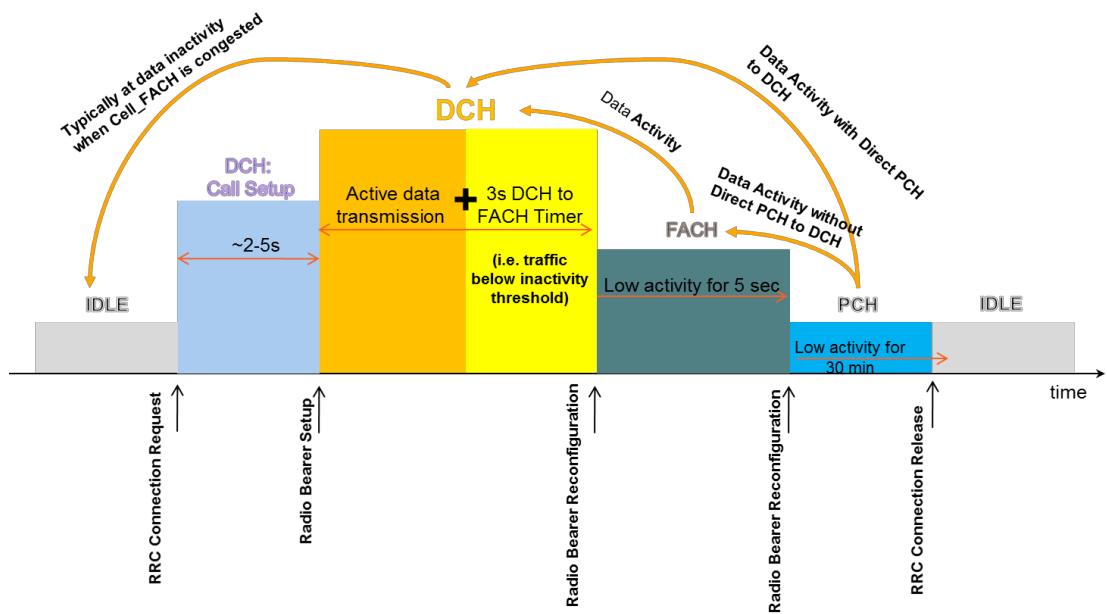

Figure 3: Different states transitions

As a result, the Cell-PCH state is expected to improve the overall network resources and KPIs. In order to practically evaluate the Cell-PCH state, we enabled it in a commercial UMTS network and we compared the main network KPIs before and after enabling the Cell-PCH state. The UMTS network consists of 5 RNCs with aggregated IuB throughput of ~6.11Gbps and two SGSN nodes with active-active geo-redundancy configuration and total Gn capacity of ~6.11Gbps. Initially the network is only deployed with RRC idle mode and both Cell-DCH and Cell-FACH. We have then enabled Cell-PCH and monitored the network KPIs. Table 1 summaries the average network improvements and degradations in main KPIs after enabling the Cell-PCH state.

Table 1 indicates that Cell-PCH reduced the RNC signaling load by 5% and DL/UL channel element (CE) utilization by 19% and 16%, respectively. The CE is the main resource element in the baseband processing board of the Node-B. The Node-B dimensioning is based on CE and the CS/PS services consume from these resources. Also, the signaling load of the Iub interface between the Node-B and the RNC has been reduced by 14% after enabling the Cell-PCH state. Moreover, RRC attempts has been significantly reduced by 28% as the UE will stay most of the time in the Cell-PCH state instead of the RRC idle mode. Other merits include 6% increase in HSDPA traffic and 3% reduction in RNC paging level. The main disadvantage is the increase in the PS call drop rate (CDR). However, it is normal to have this increase as the total number of UEs staying in connected mode has been also increased. Therefore, the PS CDR shall be updated to consider this change. The other aspects of FACH utilization increase are discussed in the last section. Finally, the PS data paging utilization has reduced by 9%. This is because the transition from Cell-DCH to Cell-PCH (i.e. D2P) happens through Cell-FACH state in this implementation. Since, the Cell-FACH is a connected state; the paging mechanism is different from the idle state because the RNC knows the UE serving cell. The observed paging reduction comes from the fact that the UE does not move to idle state as frequent as before, which improves the repeated paging at the cell level within the RNC. However, it is worth to note that enabling the direct D2P transition upon no data activity without going through Cell-FACH state can improve FACH channel congestion for the low-activity users, but it may increase the state transitions if data is inactive for very short periods. Therefore, optimizing the state transition timers and thresholds are important in this scenario. The direct D2P is a new feature that deserves to be tested and validated with existing handsets.

Table 1: Cell-PCH state deployment results

| KPIs | Change in KPIs | Comments |
|---|---|---|
| RNC Signalling Processing Boards load | -5% | Major reduction in signaling load on RNC, SGSN, and Node-B |
| RNC/SGSN Interfaces Boards Load | -1 % | |
| Channel Element Utilization (DL) | -19% | |
| Channel Element Utilization (UL) | -16% | |
| Iub Interface signaling load | -14% | |
| RRC Attempts | -28% | |
| DCH to HSPA (DtH) Success Rate | +42% | Increase in HSDPA traffic |
| HSPA to DCH (HtD) Success Rate | +3% | To conduct R99 voice call |
| HSPDA Traffic | +6% | Due to DtH increase |
| Paging (Idle, PCH) | -3% | |
| CS Drop Call Rate | - | No change |
| PS Drop Call Rate | +0.2% | Due to Cell-PCH state |
| FACH Utilization | +1.5% | FACH is expanded |
| PS data Paging Utilization | -9% | Due to FACH state |

### III. Smartphone Profiling and Signaling load

In order to quantify the impact of smartphone on commercial UMTS networks, we have conducted a device profiling analysis on a commercial UMTS network. The analyzed network has two adjacent 3G carriers at 2100MHz with approximately 3000 sites. All the cells are deployed with DC-HSDPA to support peak throughput of 42Mbps. The total number of active and unique subscribers in this network is approximately 5 million. The profiling is conducted by analyzing the RNC and PS core logs for two weeks. However, we considered a weekend day due to the volume of the data and also to demonstrate the loaded scenario for signaling analysis.

Figure 4 illustrates the terminals penetration. The data cards and 3G Routers penetrations are relatively low at 0.1% for each. The tablets came next at 0.9%. In this exercise, the feature phones represent 50% of the customers' base. The feature phone is a low-end device such as 2G/EDGE only devices or the early versions of 3G devices such as R99 and early version of HSDPA. The smartphones users represent 47.5% where 15% of which are with legacy operating system (OS) smartphones (referred to as S). The OS of the smartphones with the highest penetration (referred to as A) represent 21% of the total subscriber base while the next OS smartphones (referred to as I) is 7% followed by the last OS smartphones (referred to B) at 4%.

The terminal call setup proportion for CS and PS calls are illustrated in Figure 5 and 6, respectively. As shown in the figures and despite the 47.5% penetration of the smartphones, they consume 88% and 89% of CS and PS call setup respectively. The mapping between smartphones OS penetrations in Figure 4 and the corresponding CS call setup proportion in Figure 5 is almost logical. However, for PS call attempts in Figure 6, the smartphones with OS A, OS I, and OS B consumes relatively higher compared to the legacy OS S smartphones i.e., the penetration of OS S is 15% while the corresponding PS call setup proportion is only 3.5%. This exercise was conducted on the same network in the previous section with the Cell-PCH enabled but fast dormancy was not enabled yet.

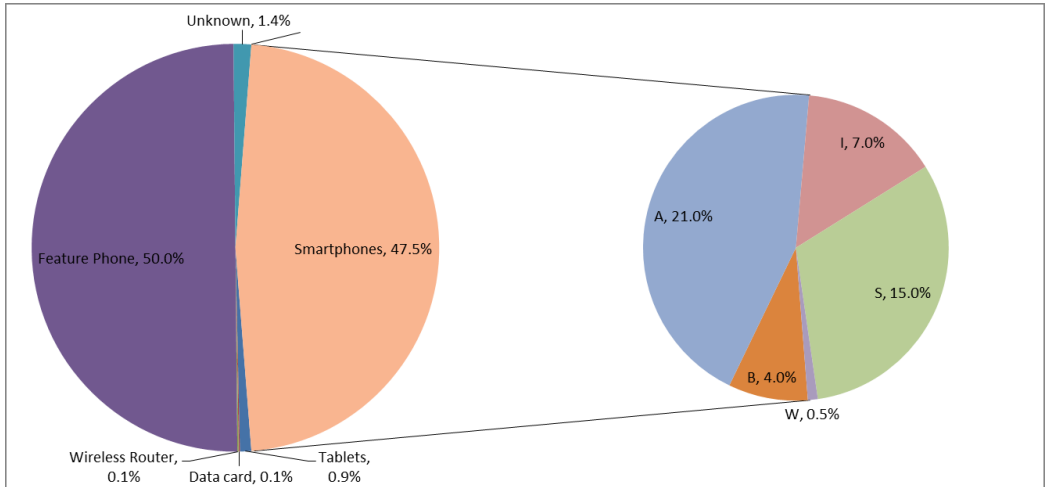

Figure 4: User Terminal Type Proportion of total attached terminals.

The most surprising behavior is the tablets that consume 10% of the PS call setup while their penetration is only 0.9%. This behavior reflects the nature of these devices and how the users use the tablets. This is an indication on the signaling storm that we will face in the near future when the tablet penetration increased to 10% for example.

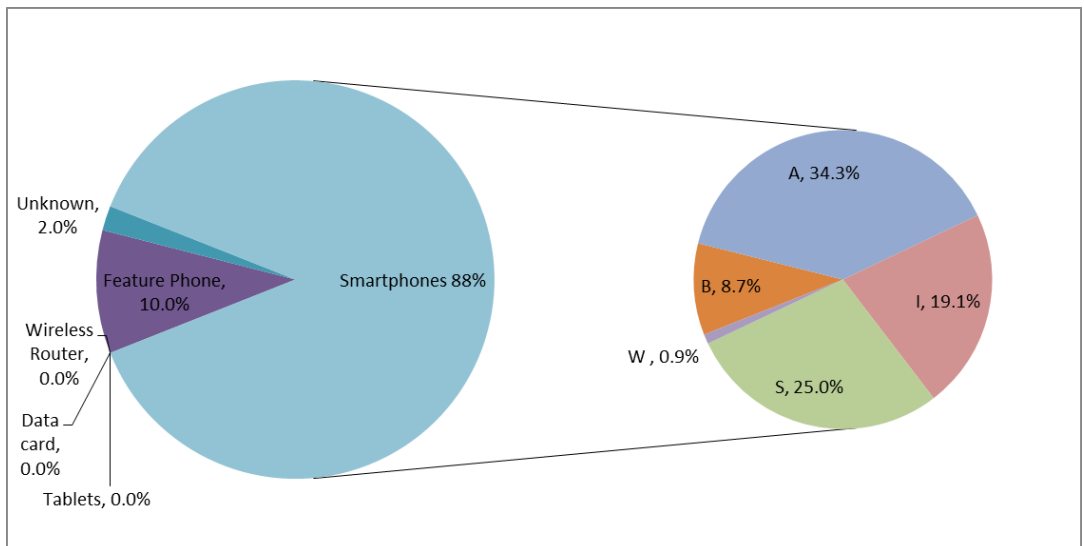

Figure 5: Terminals call setup proportion for CS calls

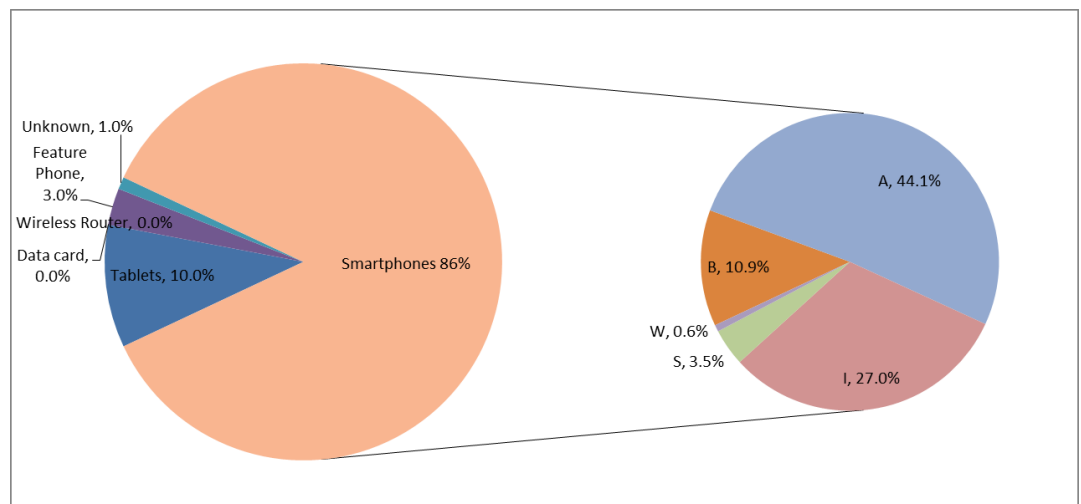

Figure 6: Terminals call setup proportion for PS calls

In addition to the call setup proportion, the Inter-RAT (Radio Access Technology) proportion is presented in Figure 7. Also, the smartphone consumes 89% of the total Inter-RAT. The Inter-RAT signaling occupies approximately 10% of the total signaling load and may cause high call drop rate. As shown in Figure 7, the smartphones with S OS causes highest Inter_RAT compared to their penetration. This is due to the 3G-receiver design in these early devices that do not support advanced interference cancellation. Therefore, it is expected that these devices will frequently handover to 2G more than the high profile smartphones with advanced interference cancellation receivers that add 3dB gain to the received signal level. The smartphones with OS I exhibit the best performance in terms of Inter-RAT proportion. This may be due to enhanced receiver/antenna design, which allows these smartphones to remain for a longer period on the 3G network compared with other high-profile smartphones.

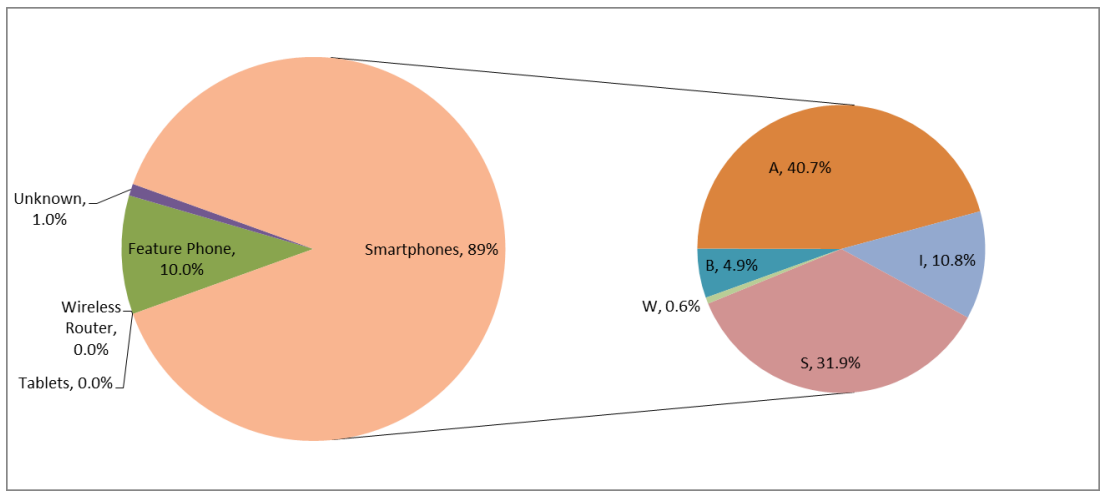

Figure 7: Inter-RAT proportion

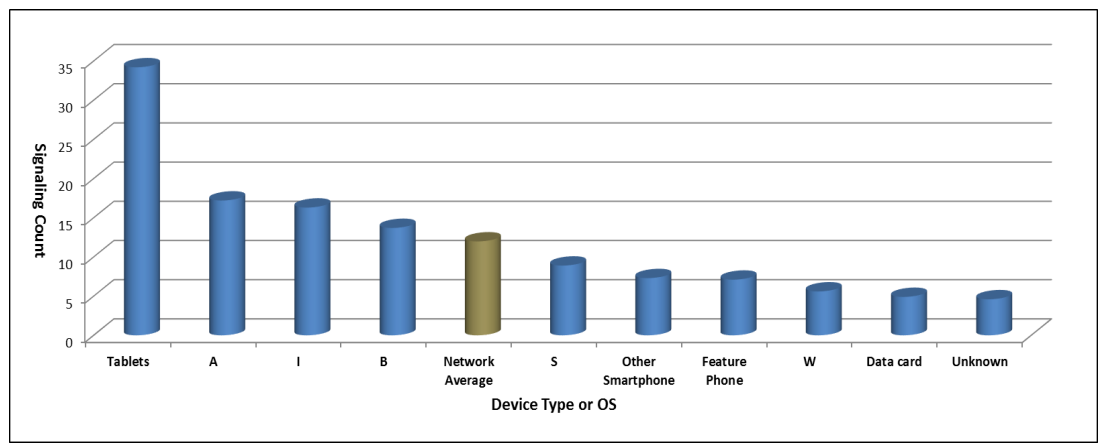

Figure 8: Signalling count by terminals at busy hour on a weekend day

Figure 8 provides the normalized average signalling count for different device types at busy hour of a weekend day. Top 3 signaling model are Tablets, smartphones with OS A and smartphones with OS I. Tablets and smartphones with OS A, OS I, and OS B consumes signaling load higher than the network average value. Tablets consumes signaling twofold the highest smartphone i.e., OS A and almost threefold the OS S smartphones. This is due to the nature of the users adopted these high profile devices and also high profile smartphones stay for a longer period on the 3G network as illustrated in figure 7. Figure 8 clearly indicates that smartphone penetration increase is a major bottleneck for the network in terms of signaling and smart approaches are mandatory to tackle the expected signaling load.

Figure 9 and Figure 10 demonstrate, respectively, the signalling count in millions and the RNC aggregated throughput in Gbps at a weekend day i.e., 24 hours. The camping users count is illustrated in both figures. The figures indicate that the aggregate user peak throughput is 5Gbps at busy hour (BH) engenders ~22M signalling at the BH which is 18:00. This data was captured in the same network with Cell-PCH enabled while the E-FD was not implemented yet. Therefore, this network needs RNCs that can manage total throughput of 5 Gbps at 80% load i.e., 6 Gbps (current capacity is 6.11Gbps). Therefore, three high capacity RNCs can manage the network in terms of throughput requirement each with 2.5Gbps. However, the network has 5 RNCs in order to be able to manage the signalling load of 22M at BH. Therefore, each RNC is managing on average 4.4 Millions. Moreover, if the tablets penetration is increased to 10%, i.e., 100% increase in the signaling load, additional 6 RNCs is needed to mange only the signaling load due to 10% increase in tablets penetration.

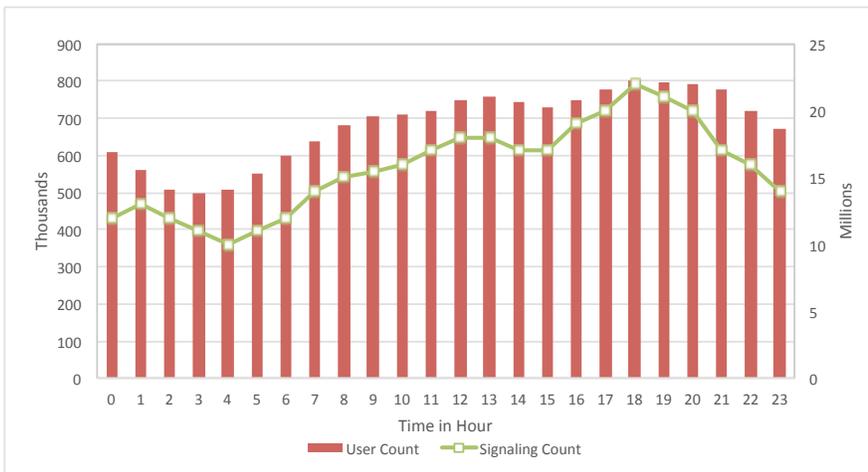

Figure 9: Signalling count and number of users at different hour during a weekend day.

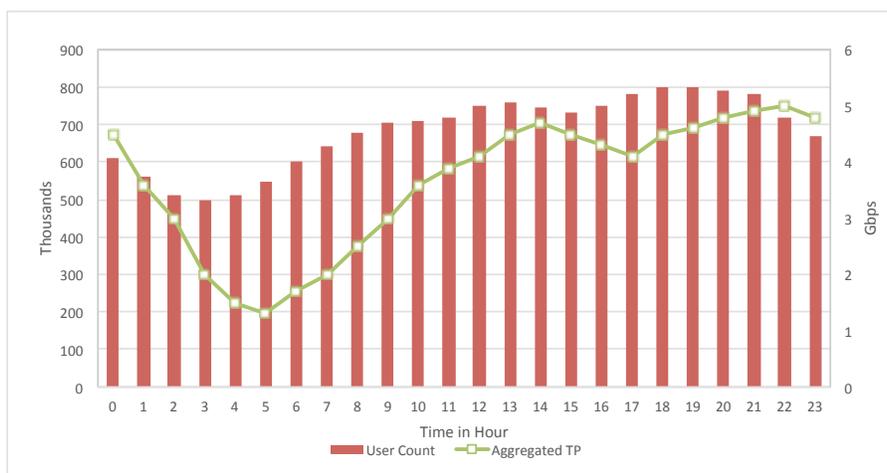

Figure 10: Aggregated throughput and number of users at different hour during a weekend day.

### IV. Fast Dormancy Mechanisms

Smartphones device manufacturers initially adopted the fast dormancy as a proprietary solution. The main driver was to release the dedicated connection as fast as possible right after a period of data inactivity. The benefit intended from this solution is to extend the UE battery backup time. This type of dormancy is referred to as legacy fast dormancy or Pre-R8 fast dormancy.

Pre-R8 fast dormancy is not standardized and its mechanism is generally implemented as proprietary by each device with different inactivity timers and different implementations techniques. These different mechanisms however had one common behavior that the UE detects a period of data inactivity and then sends RRC message Signaling Connection Release Indicator (SCRI) without any pre-defined cause. This type of UE's proprietary fast dormancy led to an increase in the signaling load on the network side. The reason is that the SCRI sent by UE requests the RNC to release the PS connection without containing the specific cause of the release, as SCRI does not mandate UEs to send a cause as per 3GPP versions prior to Release 8. Early RNC software releases could understand this SCRI as a protocol error since the UE request an RRC connection release without cause. Hence, the RNC transfers the UE quickly to idle mode. Upon any new PS data activity, the UE and RNC need to setup a new PS data connection, which increases the load on the signaling control plane.

Therefore, 3GPP have introduced a new feature to the specifications referred to as Enhanced Fast Dormancy (E-FD), also known as Rel-8 Fast Dormancy or Network-controlled Fast Dormancy. UEs complying with 3GPP Release 8 support the fast dormancy function, as defined in the change request 3GPP TS 25.331 CR3483 [16] of [17].

The Rel-8 fast dormancy mechanism is standardized as the UE sends a SCRI message to the RNC with cause set to "UE Requested PS Data session end" [16], [18]. The RNC can identify this type of UE by the SCRI message with this information and can move the UE to other RRC connected states such as Cell-FACH or Cell-PCH (i.e., D2P). As a result, the operators can see the improvement of the deployed Cell-PCH state when the fast dormancy is under the full control of the RNC.

Figure 11 describes the comparison between three PS data state transitions. The first mechanism is without any type of fast dormancy. When the PS data activity is low for a period of time, the RNC performs a state transition from Cell-DCH to other supported states such as Cell-FACH in this example. This mechanism is controlled by network timers and thresholds and can also be assisted by the UE's Traffic Volume Measurement (TVM) feedback for data activity and buffer status on the uplink. The second mechanism describes the legacy fast dormancy implementation, exactly as explained above. The third mechanism is the E-FD mechanism. In addition to introducing a cause to the SCRI, E-FD also uses a timer mechanism called T323. This timer can inhibit excessive amount of signaling and thus we control the signaling occurrences. This timer is configurable in a range between 0 and 120 seconds and is started once the UE sends E-FD SCRI and no further SCRI can be sent by UE as long as the timer is running. With this timer, the RNC can control which state to move the UE after multiple expiry. Thus, the network can move the UE first to Cell-FACH followed by Cell-PCH. In Cell-PCH, the RNC depends on a longer inactivity timer before moving the UE down to RRC idle mode.

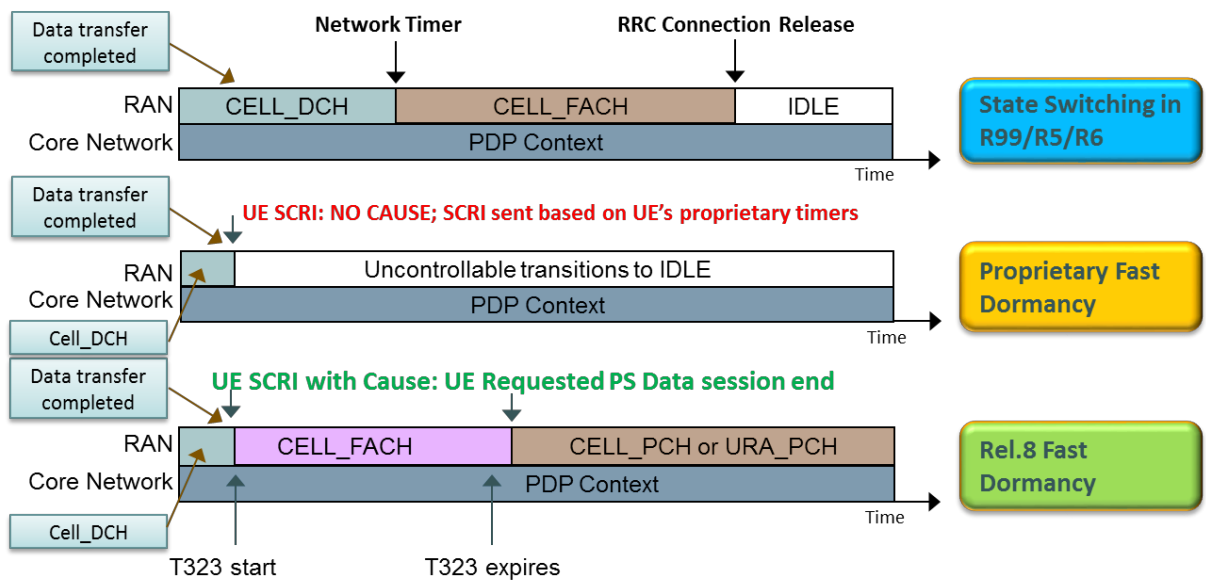

Figure 11: Mechanisms of no Fast Dormancy, legacy Fast Dormancy, and Enhanced Fast Dormancy

The different types of commercial devices support different types of fast dormancy implementations. For the devices supporting E-FD, the RNC can efficiently handle them with the known behavior of SCRI with a specific cause. However, with devices that support only the legacy fast dormancy, which are the pre-R8 devices, we need to have smart approaches to handle their behavior and prohibit their transition to the idle mode, generating excessive signaling load. Hence, one of the approaches is to handle these known devices by their IMEI (International Mobile Equipment Identity). For those devices, the RNC can follow the exact state transition behavior as the E-FD and control their behavior in terms of signaling load. However, the RNC does not have any control over how frequent these devices reporting the SCRI and therefore the T323 timer applies only to the standardized E-FD solution.

To fully utilize the benefits of the E-FD, we must deploy multiple states including Cell-FACH and either Cell-PCH or URA-PCH. Additionally, the transition to RRC idle mode for PS services should be set to a high value after a period of inactivity. Figure 12 demonstrates the the fast dormancy procedures and expected gain in terms of signaling load and battery saving. The practical gains of fast dormancy implemented alongside Cell-FACH and Cell-PCH are discussed the next section.

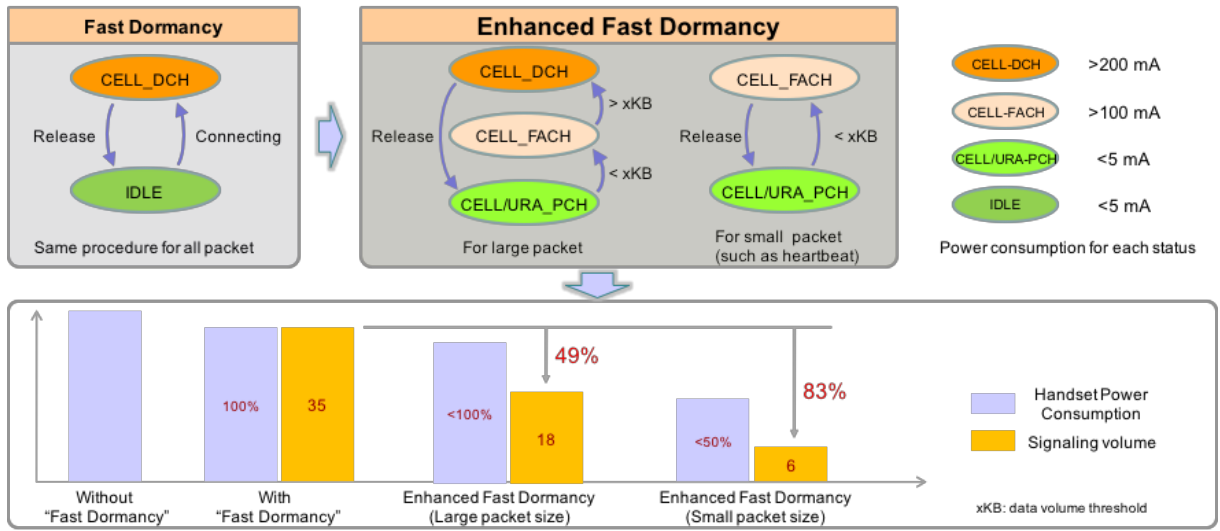

Figure 12 Enhanced Fast Dormancy expected gains

## V. Practical Results of E-FD deployment

The test results presented in this section are obtained from one RNC of a commercial UMTS network. The RNC is deployed with two 3G carriers. The results shown are extracted from the RNC-wide KPIs and for all active users present in the RNC during the monitoring period. The main KPIs analyzed in this trial are summarized in table 2. In order to evaluate the expected gains of the KPIs described in table 2, we monitored the cells in the selected RNC. The most important KPIs that have a direct impact on the signaling load are the ones related to establishment of RRC control plane connections as well as the RAB establishment attempts for PS data calls.

The KPIs of the RNC are presented for a period of two months. The E-FD is implemented on the end of day 24[th] of the month and since the KPIs are averaged over 24 hours, therefore the changes are fully reflected on day 26. As shown in figure 13, the number of RRC connections is reported on a daily basis for all cells in the RNC. After the E-FD is enabled, the number of RRC connections has decreased significantly by 59%. This improvement is achieved because the E-FD UEs will experience more state transitions to Cell-FACH and Cell-PCH rather than to RRC idle mode. When the PS data becomes active, the UE and RNC need less number of RRC signaling messages compared to data re-establishment from idle mode. Because of this gain, the RNC/NB CPUs signaling load, the UE power consumption and the transmitted power are reduced accordingly. The power consumption in Cell-PCH state is almost similar to the Idle state because we are using the same DRX cycle length i.e., paging cycle in the two states. Also, all other idle mode reselection parameters are same in the network for both idle and PCH states since the network is configured with SIB3 reselection parameters only and SIB4 was not configured. Therefore, the reduction in RRC signaling load reduces the UE power consumption as illustrated in figure 12. Moreover, the smartphones applications that are working in the background are frequently move the UE to the Cell-DCH state and hence the reduction in RRC signaling load will extended the battery life.

Table 2: RNC KPIs used to evaluate E-FD Implementation

| KPI | Definition | Expected Impact with E-FD |
| --- | --- | --- |
| RRC Connection Establishment Attempts | Number of Processed RRC Connection Requests for all Cells in RNC | Reduced as transitions between Idle and Cell-DCH are reduced |
| RAB Establishment Attempts | Number of PS Call RAB Establishment Attempts for Cells in RNC | |
| FACH, PCH and DCH Users | Number of users served in Cell-FACH and Cell-DCH states | Users in Cell-DCH reduced while Users in Cell-FACH and Cell-PCH increased, as Fast Dormancy reduces the time for a user in Cell-DCH when data is inactive |
| Cell Update Attempts | Number of Cell Update sent in Cell-FACH due to cell re-selection or Data Transmission | Increased as the number of users in Cell-FACH and Cell-PCH also increased |
| Transitions between Cell-FACH and Cell-DCH (HSPA) | Number of state transitions between Cell-FACH and HSDPA | Increased as the users are being served more in Cell-FACH state |
| PS Paging Type1 Losses | Number of PS paging attempts and losses | Reduced Paging Type1 losses as UE can still be served in FACH for more time which reduces the need to page users when DL data is available |

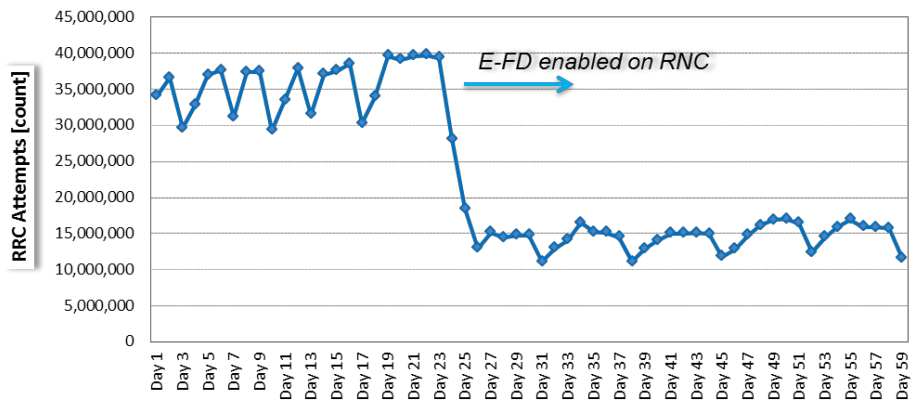

Figure 13: RRC Connection Establishment Attempt KPI with and without E-FD

The number of PS Call RAB establishment attempts for all cells in the RNC has decreased by 82% as demonstrated in figure 14. The reason for this improvement is related to the mechanism of E-FD, which triggers the users to Cell-FACH or Cell-PCH where the PS RAB is kept active instead of being released.

As observed so far, the E-FD is forcing the users to move to Cell-FACH and Cell-PCH more often compared to the case without E-FD in the same RNC. Therefore, it is expected to see an increased number of users being served in Cell-FACH and Cell-PCH states. This behavior is demonstrated in figure 15. The number of users in Cell-FACH has increased 116% and the umber of users in Cell-DCH has increased by 246%. The number of users in DCH state has reduced by 15% as shown in figure 15. The gain of reducing DCH utilization is very significant to the system performance where relevant dedicated channels are released. It evident from figure 14 that the users in Cell-PCH have increased more than those in Cell-FACH because the E-FD is triggered when data activity is really low and hence there is no reason for the RNC to keep inactive users in Cell-FACH for longer period. This is illustrated in figure 16 where the transition from Cell-FACH state to Cell-PCH state has increased when E-FD is enabled.

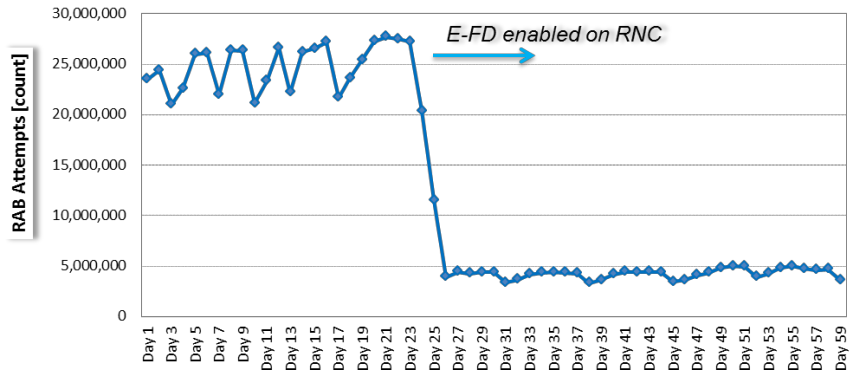

Figure 14: PS RAB Establishment Attempt KPI with and without E-FD

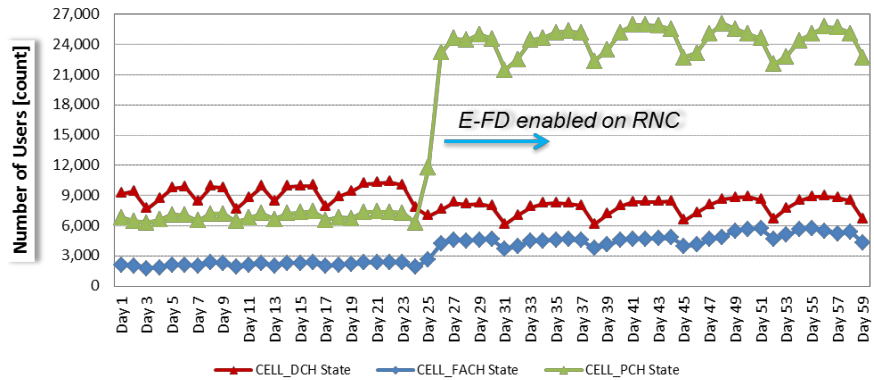

Figure 15: Number of users in Cell-FACH, Cell-DCH and Cell-PCH states with and without E-FD

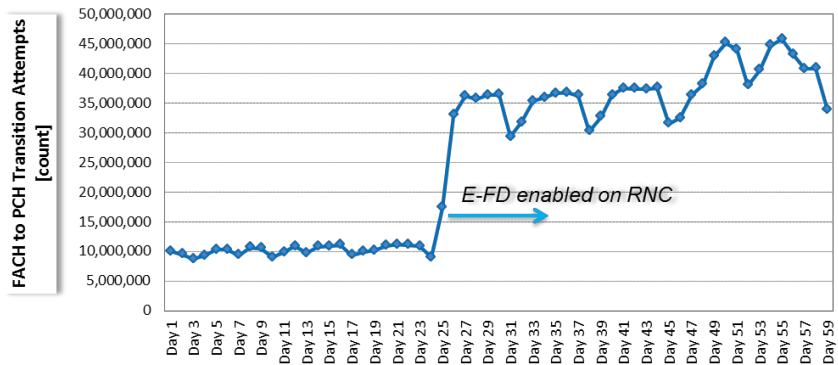

Figure 16: Cell-FACH to Cell-PCH Attempts KPI with and without E-FD

As the number of users has increased in Cell-FACH and Cell-PCH states, then it is expected to see more Cell update messages in the RNC. Cell update is one of few RRC signaling messages sent in these states to keep the UE synchronized with the RNC. Cell updates are sent by UE whenever the UE re-select to a new serving cell during mobility or when data becomes active in Cell-PCH state. Upon data activity in Cell-PCH, the UE needs to inform the network to switch it back to Cell-FACH or DCH states in order to complete the PS data activity. Figure 17 demonstrates that the number of Cell updates sent by UE has increased by 226% after deploying E-FD in the RNC. The increase in Cell updates may cause several degradations to other KPIs in the network including voice call accessibility. This is discussed in details in the next section.

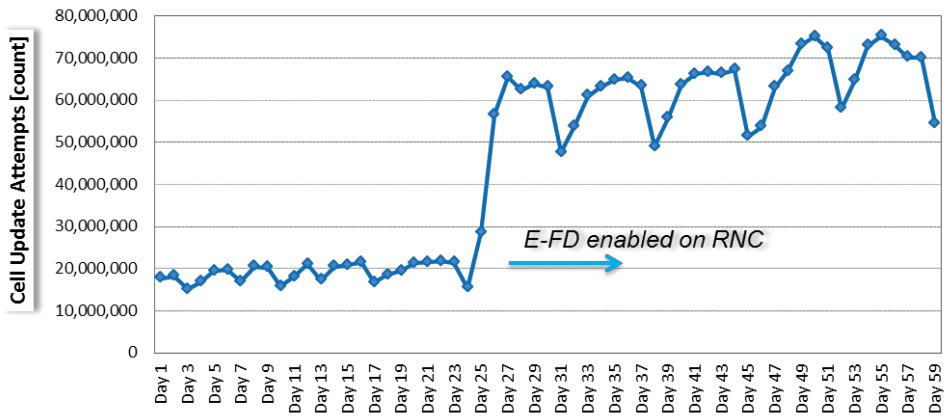

Figure 17: Cell Update Attempt KPI with and without E-FD

On the other hand, the state transition between Cell-FACH and HSDPA (i.e. Cell-DCH state where HSDPA service is active) has also increased by 197%, as shown in figure 18. This increase also requires careful optimization to avoid any degradation to other KPIs and explained in next section as well.

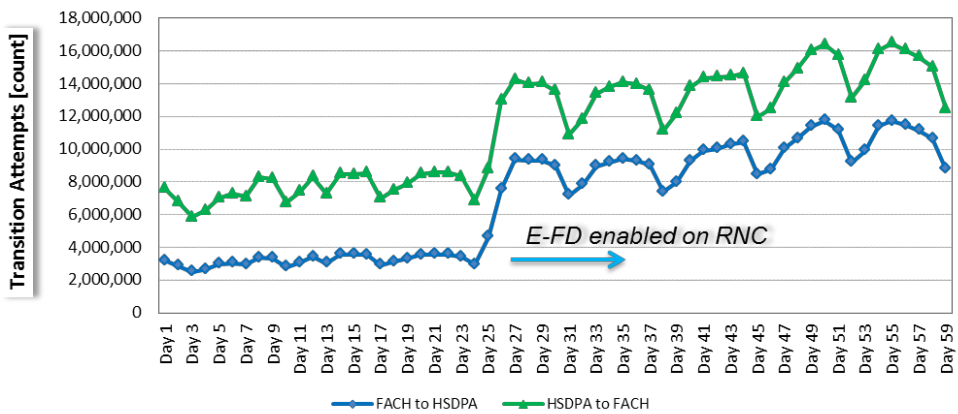

Figure 18: Cell-FACH to HSDPA, and HSDPA to Cell-FACH attempts with and without E-FD

It is important now to evaluate the paging congestion when E-FD is enabled. Typically, there are two ways to start the PS data activity. During the idle mode if the UE detects data in its uplink buffer, it will send a new RRC connection to establish the PS domain. On the other side, if the RNC detects data in the downlink buffer and the UE is in idle mode, the network will page the UE using Paging Type 1 message for the PS domain and then establish an RRC connection. The same procedure applies for UEs in Cell-PCH for the PS connection re-establishment. The difference is that in the Cell-PCH state and after the UE receives the Paging Type 1, it will directly send Cell update and the network moves the UE to Cell-FACH or Cell-DCH in order to continue with the data transfer. With E-FD, the UEs are well controlled to be in FACH or PCH depending on the amount of data in the buffer. However, with legacy fast dormancy or without any fast dormancy, the UE could be served more in idle mode with very frequent paging attempts to re-establish the PS connection.

As illustrated in figure 19, the E-FD offers significant gain on the number of paging attempts (reduced by 65%) and the paging losses (reduced by 99%). This has a positive impact on the network performance as the reduction in paging attempts leads less congestion in the core network and network accessibility. The paging losses has reduced because the overall signaling load has decreased and consequently the core network resources have improved to grant more paging to the UEs in the system.

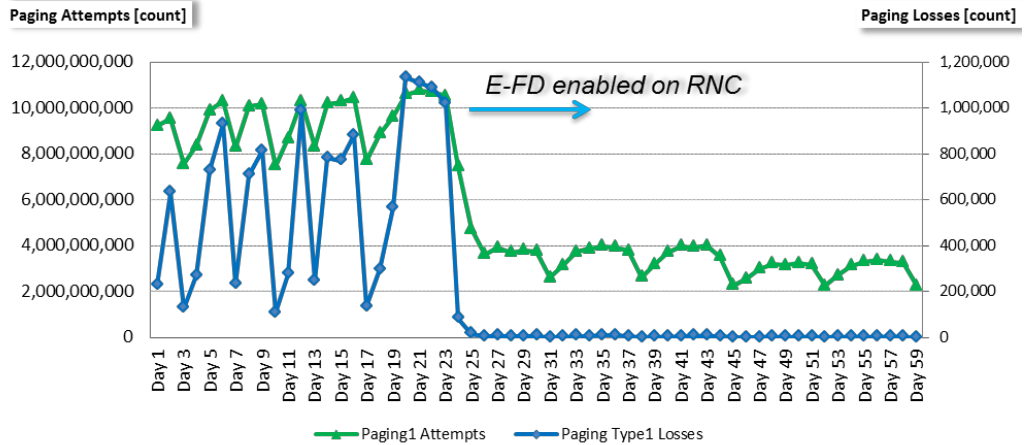

Figure 19: Paging Attempts and Paging Losses KPIs with and without E-FD

In general, the main KPIs related to signaling load are significantly improved after enabling the E-FD along with Cell-PCH. Table 3 summarizes the overall impact on all related KPIs. Figure 20 demonstrates the CPU load on the impacted network elements after E-FD activation. The saving in base band channel elements increases the Node-B capacity and reduces the signaling load on the Node-B CPU. Also, the call drop rate for PS R99 traffic has significantly improved by 44%. Finally, the mean transmit power in dBm for one site is provided in figure 21. The figure demonstrates maximum of 0.5dBm reduction in transmit power at busy hour from 18:00 to 19:00. The reduction in total transmit power reduces the DL interference and also increase the overall network capacity.

Table 3: The impact of E-FD on the relevant KPIs

| KPI | Average Value Before E-FD | Average Value After EFD | Impact on KPIs | Comments |
|---|---|---|---|---|
| **Number of UEs in CELL-DCH State** | 9,299 | 7,902 | -15% | Sum of # of Users served in Cell-DCH reduced by 15% |
| **Number of UEs in CELL-FACH State** | 2,160 | 4,665 | 116% | Sum of # of Users served in Cell-FACH increased by 116%. FACH expansion is needed |
| **Number of UEs in CELL-PCH State** | 6,950 | 24,073 | 246% | Sum of # of Users served in Cell-PCH increased by 246% due to dormant Traffic. Positive Impact on UE battery |
| **Call Drop Rate for PS R99 Traffic** | 0.221 | 0.123 | -44% | Reduction in # of Users in Cell-DCH improves resources to serve R99 traffic |
| **Call Drop Rate for HSDPA Traffic** | 0.222 | 0.341 | 54% | Degraded KPI, due to increase in HSDPA traffic and frequent state transitions |
| **RRC Connection Establishment Attempts** | 35,676,287 | 14,645,168 | -59% | Significant improvement in signaling due to less RRC establishment from Idle with E-FD |
| **PS RAB Establishment Attempts** | 24,857,711 | 4,453,183 | -82% | Significant improvement in signaling due to less RAB establishment from Idle with E-FD |
| **Cell Update Attempts** | 19,176,219 | 62,530,844 | 226% | Increased Cell Update due to increased # of users in Cell-FACH and Cell-PCH. Requires further optimization |
| **Cell-FACH to HSDPA Transition Attempts** | 3,192,577 | 9,481,118 | 197% | Increased state transitions can introduce more call drops, depending on the Multi-carrier Strategy |
| **HSDPA to Cell-FACH Attempts** | 7,649,678 | 13,849,229 | 81% | |
| **PS Paging Type1 Losses** | 611,886 | 3,415 | -99% | The paging losses have reduced because the overall signaling load has decreased |

| PS Paging Type1 Attempts | 9,473,758,526 | 3,359,632,080 | -65% | Reduction in Paging Attempts because of UE is being served in Cell-FACH before transition to Cell-PCH while inactive due to E-FD |

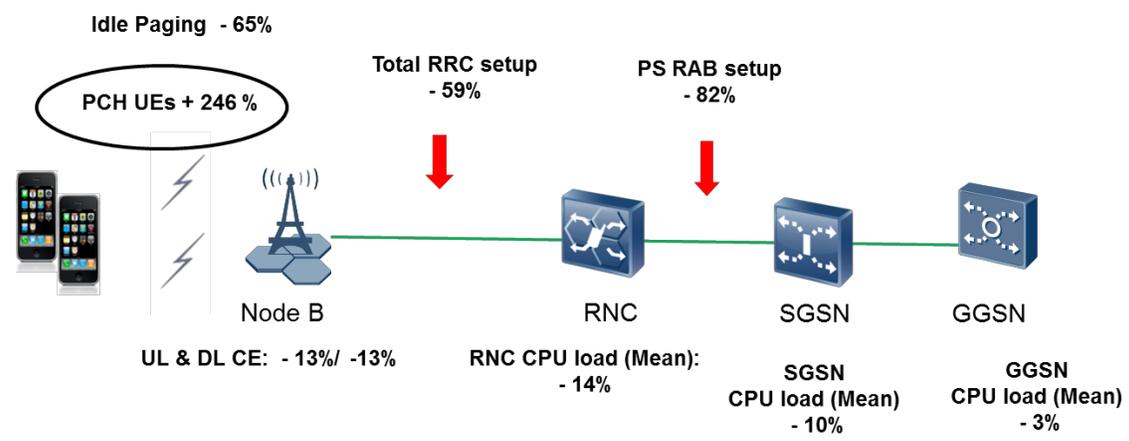

Figure 20: Access and Core Networks CPU load impact after E-FD activation on one RNC

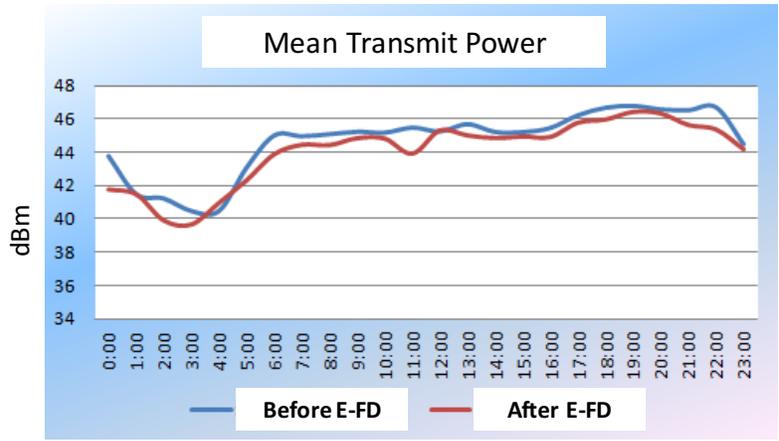

Figure 21: Mean transmit power for a congested site in dBm.

## VI. Cell-PCH and Cell-FACH Optimization

To avoid any further degradation to KPIs such as circuit switch (CS) voice accessibility, there is still a room for further improvement to state transitions and congestion relief in Cell-FACH state. In the previous section, we demonstrated the gains achieved from E-FD to the overall system KPIs. However, there are several areas where E-FD can introduce additional issues that need further optimization. The two main areas of improvements after E-FD deployment are optimizing the Cell-FACH congestion and the CS voice call setup accessibility from Cell-FACH and Cell-PCH states.

The main drawback observed after deploying the E-FD is the increase in the call drop rate. Figure 22 presents a slight increase in the HSDPA call drop rate after deploying the E-FD. The typical reasons for such call drop increase are the frequent state transitions between FACH/PCH/DCH states as well as the lack of power control or increased congestion in FACH state. Other areas of improvements include reduce the amount of Cell update triggered by frequent mobility cell re-selection that can add into call dropped scenarios.

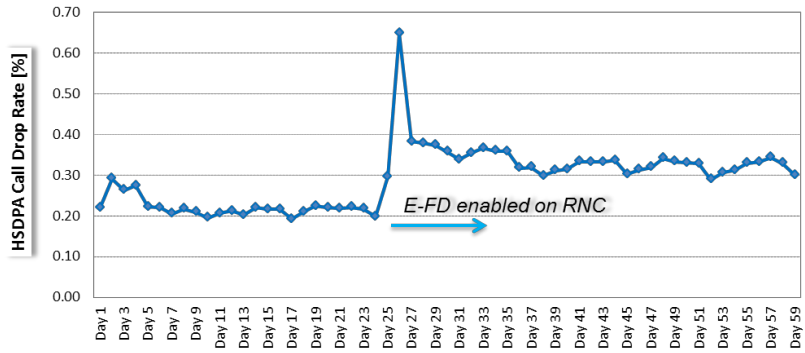

Figure 22: Service Call Drop Rate with and without E-FD

*a)* **Cell-FACH Congestion**

From the results shown in previous section, we illustrated that the E-FD had a significant gain on network KPIs. However, this comes on the expense of degradations to other services including degradation in FACH performance due to loading and number of users that can be supported in FACH.

Fast dormancy introduces more utilization to FACH channel and therefore FACH congestion is expected. This is evident in the increase in the number of users served in FACH after E-FD is enabled, as shown in figure 15. Therefore, smart solutions are needed to handle this behavior. The UEs served in Cell-FACH state are scheduled on a common channel on the downlink. This means that the bandwidth of the physical layer channel is limited. Practically, the maximum PS data speed of physical layer channel in Cell-FACH is limited to 36 kbps, for the traffic part of this channel. This effectively means that if the UE is served for a longer duration in Cell-FACH state with E-FD, then it will transmit more data in this state. With the higher number of users in Cell-FACH, more congestion is expected. To mitigate this congestion, it is recommended to increase the data rate of the physical layer channel carrying the Cell-FACH. This is practically achieved by increasing the number of Transport Blocks (TB) to 2 instead of 1, which increases the throughput to 72 kbps. This increase in the data rate will allow the users to complete the data transfer faster and move to the next inactive state (i.e., Cell-PCH). Figure 23 illustrates the congestion of the DTCHs (dedicated traffic channel) carried on the FACH with one TB and with two TB after the FACH expansion of a certain cell. The illustrated KPI provides the congestion durations in seconds of the DTCHs carried on the the FACH. As shown in the figure, the congestion of the DTCHs carried on the FACH is reduced significantly after the expansion of the FACH. The KPI counts how many seconds during a predefined period the cell experienced DTCH congestion (messages in the buffer of this channel can not be sent). The congestion on the DTCH is cleared when the messages saved in the buffer of the DTCH can be sent out.

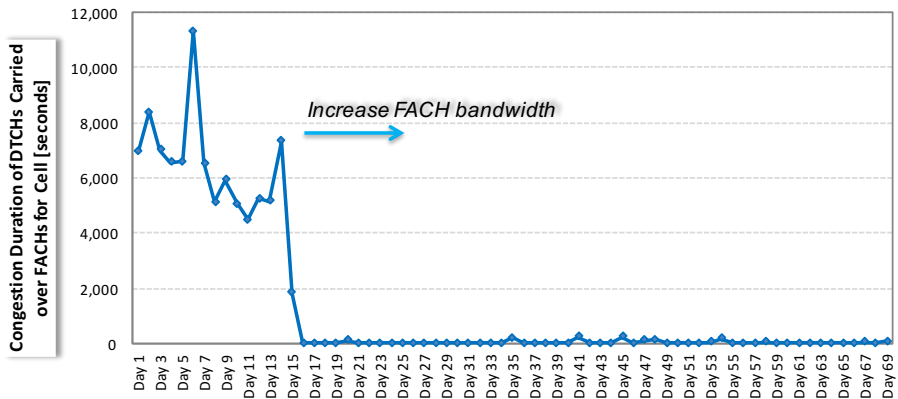

Figure 23: DTCHs carried on FACH congestion before and after expanding FACH channel bandwidth

### b) *Impact on Voice User Accessibility*

When E-FD is enabled, the UE are going to be served more in FACH or PCH compared to the UE when served in Idle state without E-FD. Hence, CS failures are higher in Cell-PCH and Cell-FACH states due to conditions of re-selection and lack of power control. This requires an additional step of optimization to the CS voice call performance.

Prior to E-FD deployment, the users are mostly initiating CS voice calls from idle mode. Most of the optimization process has been targeting improvements to CS accessibility and call setup time from idle mode. This is typically conducted by tuning the cell re-selection parameters and timers related to CS setup from idle state. However, with E-FD, the re-selection process is required to be reviewed as most of the CS calls will be initiated from either Cell-FACH or Cell-PCH.

During idle mode, both FACH/PCH mobility depend on cell re-selection and not handover as in Cell-DCH, the re-selection procedure can still be different. In idle mode, when the UE moves to a new cell, it is not necessarily to send any RRC message to inform the network of its new cell-level location. However, in FACH/PCH states, the UE needs to initiate cell update procedure to indicate its new serving cell. Additionally, as FACH and PCH are still connected mode states, the security mode procedure needs to be handled when a CS call is initiated. Therefore, it is expected to have CS setup failure cases especially when a collision between Cell update and CS setup call initiation occurred.

3GPP provides the options to distinguish between cell re-selection parameters in idle mode and in FACH/PCH states. In idle mode, all re-selection parameters are broadcast in SIB-3. For FACH and PCH states, the same parameters can be modified if the cell broadcasting SIB-4. The main issues observed when cell re-selections frequently occur. Therefore, it is important to reduce the number of cell re-selection in FACH and PCH states which can reduce the number of collision between Cell update and CS voice call setup. A reduction in cell update is also beneficial to the signaling load, which is observed to increase with E-FD as shown in figure 16.

There are two ways to reduce the frequency of cell re-selection in FACH. This can be done by using SIB-4 that overrides the Time to Re-selection ($T_{resel}$) parameter, or by using SIB-4 with a new $T_{resel}$ for FACH and a different one for PCH state (known as $T_{resel,fach}$ and $T_{resel,pch}$). The first method will utilize the same $T_{resel}$ value for all connected mode states, and the second one can introduce more distinction to each connected mode state: FACH or PCH states.

As shown in figure 2, several types of state transitions are allowed. One of the transitions that are not widely used is a direct move from PCH to DCH states. If this direct transition is not enabled by the network operator, then any CS or PS call establishment from PCH states is needed to go through Cell-FACH and then to Cell-DCH. This effectively means that the call setup will be handled within Cell-FACH first before the transition to Cell-DCH at the RAB setup stage. This can in return jeopardize the call accessibility especially in cases of FACH congestion or when cell update procedure conflict with the call setup requests. Therefore, introducing the direct PCH to DCH (P2D) transition is recommended which triggers the UE to temporarily move to FACH to send the cell update, followed by Cell Update Confirm that directs the UE to Cell-DCH without having to setup the call on Cell-FACH. However, the main impact of direct P2D transition is a slightly higher call setup time. The call setup time slightly increases because the signaling speed from PCH to DCH will be limited to 3.4 kbps where the call setup from idle mode is carried over signaling with a rate of 13.6 kbps. Typical average call setup time from different RRC states is provided in table 4.

Table 4: CS Voice Call Setup from different RRC States

| Starting RRC State | Average Call Setup Time |
|---|---|
| CS Setup from Cell-FACH | 2171 ms |
| CS Setup from IDLE | 2109 ms |
| CS Setup from Cell-PCH | 3297 ms |

## VII. Conclusions

In this paper, the best practices for managing smartphone signaling load are provided. We have analyzed two major features; Cell-PCH and E-FD to mitigate the smartphone signalling. The presented performance analysis of both features indicates major improvement in the major KPIs and significant relief in the signaling load. The deployment of the presented features reduce the smartphone signaling load, relief the network resources, reduce transmitted power and enhance users experience by reducing the drop call and extending battery backup time. We have provided detailed analysis for the impacted KPIs based on a commercial UMTS network. Furthermore, additional optimization techniques are provided to mitigate the drawbacks of the E-FD. The presented results can be used to benchmark the features performance. Future work includes analyzing other advanced features that relief signalling congestion and extends battery backup time such as CPC and Enhanced Cell-FACH with DRX.

**Ayman Elnashar** received the B.S. degree in electrical engineering from Alexandria University, Alexandria, Egypt, in 1995 and the M.Sc. and Ph.D. degrees in electrical communications engineering from Mansoura University, Mansoura, Egypt, in 1999 and 2005, respectively. He has more than 20 years of experience in telecoms industry. He was part of three major start-up telecom operators in MENA region (Mobinil/Egypt, Mobily/KSA, and du/UAE) and held key leadership positions. Currently, he is Head of Core and Cloud Planning with the Emirates Integrated Telecommunications Co. "du", UAE. He is responsible for strategy and innovation, design and planning, and performance and optimization of core networks and cloud infrastructure for mobile, fixed and IT. Prior to this assignment, he was Sr. Director of Wireless Networks, Terminals and IoT. He is the founder of the Terminal Innovation Lab for end-to-end testing, validation, and benchmarking of mobile terminals. He managed and directed the evolution, evaluation, and introduction of du mobile broadband HSPA+/LTE networks. Prior to this, he was with Mobily, Saudi Arabia, from June 2005 to Jan 2008 as Head of Projects. He played key role in contributing to the success of the mobile broadband network of Mobily/KSA. From March 2000 to June 2005, he was with Orange, Egypt where he was responsible for the operation & maintenance.

He published 20+ papers in wireless communications in highly ranked journals. He is the author of "Design, Deployment, and Performance of 4G-LTE Networks: A Practical Approach", Wiley May 2014. His research interests include practical performance analysis of cellular systems, mobile networks planning, design, and optimization, digital signal processing for wireless communications, multiuser detection, smart antennas, massive MIMO, and robust adaptive detection and beamforming. He is currently working on SDN/NFV and network transformation towards 5G.

**Mohamed A. El-Saidny** is currently a Sr. Technical & Carrier Account Manager at MediaTek. He is leading technical expert in wireless communication systems for mobile phones, modem chipsets, and networks operators. He received the B.Sc. degree in Computer Engineering and the M.Sc. degree in Electrical Engineering from the University of Alabama in Huntsville, USA in 2002 and 2004, respectively. Prior to MediaTek, he worked in Qualcomm CDMA Technology, Inc. (QCT), San Diego, California, USA. He was responsible for performance evaluation and analysis of the Qualcomm UMTS system and software solutions used in user equipment. As part of his assignments, he developed and implemented system studies to optimize the performance of various UMTS algorithms. The enhancements utilize Cell re-selection, Handover, Cell Search and Paging. He designed algorithms to improve the performance of 3G systems, some of which already adopted in 3GPP. He later worked in Qualcomm Corporate Engineering Services division in Dubai, UAE. He worked on expanding the 3G/4G technologies footprints with operators, with an additional focus on user equipment and network performance as well as technical roadmaps related to the industry. Mohamed's current responsibilities at MediaTek are leading operators' accounts in Middle East, North Africa and Turkey in addition to worldwide network operators and groups in LTE commercial efforts, roadmaps and technical services. He led a key role in different first time features evaluations such as CSFB, C-DRX, IRAT, VoLTE, Carrier Aggregation and load balance techniques in LTE. As part of this role, he is focused on aligning network operators to the device and chipset roadmaps and products in both 3G and 4G. Mohamed is the author of several international IEEE journal papers and contributions to 3GPP, and an inventor of numerous patents.

**Mohamed Reda** received the B.S. degree in communications engineering from Ain Shams University, Cairo, Egypt, in 2010. He has 4 years of practical experience in telecoms industry including GSM, GPRS/EDGE, UMTS/HSPA+ technologies. Currently, he is RF Optimization Supervisor with the Egyptian Company for Mobile Services "Mobinil", Egypt. In addition, he is responsible for validating new features & KPIs' equations, also terminals testing & validation and the impact of new features on them. He has managed several large-scale projects across Mobinil including E2E Data throughput enhancement, SRAN Swap & network expansion.